\documentclass[12pt]{iopart}
\usepackage{iopams} 
\usepackage{graphicx}
\usepackage{enumerate}

\begin{document}

\title[Thermal properties of composite two-layer systems]{Thermal properties of composite two-layer systems with a fractal inclusion structure}

\author{J J Reyes-Salgado, V Dossetti, \hbox{B Bonilla-Capilla} and \hbox{J L Carrillo}}

\address{Instituto de F\'{\i}sica, Benem\'erita Universidad Aut\'onoma de Puebla, \\Apartado Postal J-48, Puebla 72570, Mexico}
\ead{dossetti@ifuap.buap.mx}

\begin{abstract}
In this work, we study the thermal transport properties of platelike composite two-layer samples made of polyester resin and magnetite inclusions. By means of photoacoustic spectroscopy and thermal relaxation, their effective thermal diffusivity and conductivity were experimentally measured. The composite layers were prepared under the action of a static magnetic field, resulting in anisotropic inclusion structures with the formation of chains of magnetite particles parallel to the faces of the layers. In one kind of bilayers, a composite layer was formed on top of a resin layer while their relative thickness was varied. These samples can be described by known models. In contrast, bilayers with the same concentration of inclusions and the same thickness on both sides, where only the angle between their inclusion structures was systematically varied, show a nontrivial behaviour of their thermal conductivity as a function of this angle. Through a lacunarity analysis, we explain the observed thermal response in terms of the complexity of the interface between the layers.
\end{abstract}


\section{Introduction}
\label{sec:intro}
The study of thermal conduction in multilayered systems is of great interest nowadays, due to the development of an increasing kind of coatings for applications that span electronic and optoelectronic devices, glues and turbines among others \cite{cla05}. On the other hand, controlling thermal conduction of composite materials by means of changes in their structure and composition is an area of current research for its potential applications \cite{geo04, goy08, liu10, bou13, nar12}. In particular, controlling the inclusion structure seems to be promising in this regard, as it allows to tune the thermal properties of the composite without having to change the component materials themselves \cite{cer14}. These kinds of studies may eventually lead to the development of intelligent materials with a real-time controllable thermal response, as it has happened with some other properties like elastic, optical and electrical to mention a few. It is worth to mention that some advances have been made in this direction with magnetic fluids, as their thermal properties can be controlled with the concentration of particles and the application of magnetic fields with different strengths. In this kind of fluids, it has been observed that the development of chains by the embedded particles greatly enhances their thermal conductivity in the direction of the chains \cite{li05, phi07, fan09, aba10, gav12}.

Regarding experimental approaches, the photoacoustic (PA) technique in combination with the thermal re\-laxa\-tion method (TRM) have proven to be reliable and useful non-destructive techniques to measure thermal properties of different kinds of materials \cite{cer14, alv94, dos02, siq08}. About the first, there has been an ongoing effort to extend the photoacoustic technique for the characterization of multilayered systems, with the purpose to determine their effective thermal diffusivity from the knowledge of the thermal properties of the layers themselves \cite{gur02, gur03, tla08, ast10, pop12}. In particular, for two-layer systems, there are some aspects to be taken into account such as the thermal thickness of the layers and their effusivity, that becomes important at the interface in relation to its thermal resistance \cite{man90, mar96, pic01}. On the other hand, the TRM has been complimentarily used for the measurement of the heat capacity of small samples \cite{hat79, gar06, val06}. Jointly applied, these techniques allow the determination of the thermal conductivity of different samples.

In this work, by means of the PA technique in combination with the TRM, we study the thermal properties of composite two-layer systems. The layers consist in a polyester resin matrix with powdered magnetite inclusions that present an anisotropic fractal structure, resulting from the application of a magnetic field in the direction parallel to the faces of the layers during their preparation. Two kinds of two-layer samples were studied. For the first, resin-composite bilayers (RCBs) were prepared, where the volume fraction of inclusions of the composite layer was varied as well as the relative thickness of the layers. For the other kind, composite-composite bilayers (CCBs) were prepared, with layers of the same thickness and inclusion concentration on each side. For these, only the angle between the inclusion structures of the layers was systematically varied. The effective thermal properties of the RCBs can be described by well established models, however, the CCBs exhibit a nontrivial behaviour in their effective thermal conductivity as a function of the angle between their anisotropic inclusion structures. Results from this work confirm our previous findings \cite{cer14}, where the increase of thermal resistance was associated with the formation of overlapping domains of resin and magnetite aggregates, in this case purposefully induced at the interface of the two-layer systems by varying the angle between the inclusion structures on its sides.

\begin{figure}[t]
\begin{center}
\includegraphics[width=0.7\textwidth]{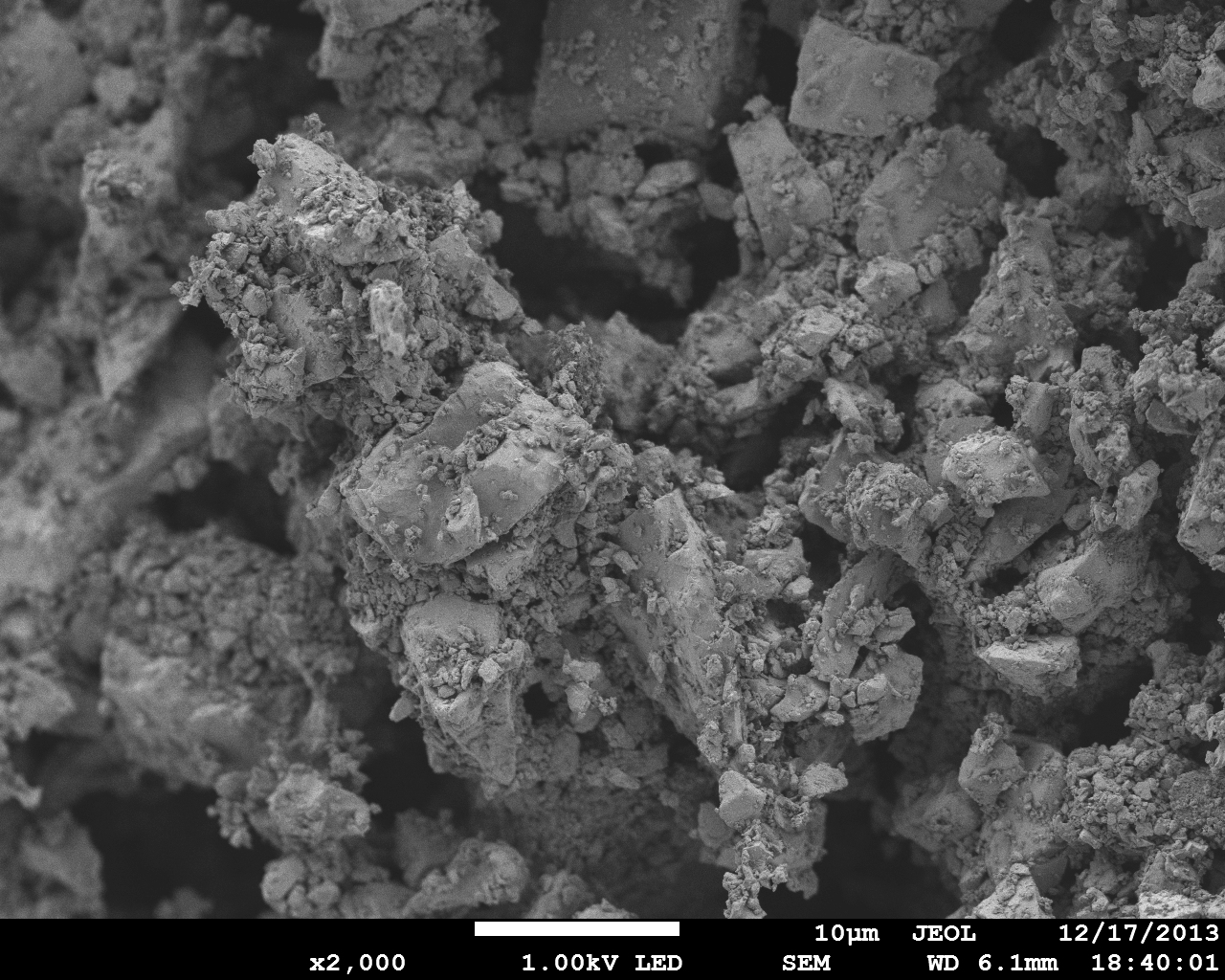}
\caption{Scanning electron micrograph of the resulting magnetite powder from the grinding and sifting process with magnification $\times 2000$. The resulting grains are very polydisperse with sizes going down to the nanometric scale as apparent from the figure.}
\label{fig:incl}
\end{center}
\end{figure}

\section{Experimental details}
\label{sec:expdet}


\subsection{Samples preparation}
\label{ssec:samprep}

First of all, we obtained the magnetite particles by grinding magnetite crystals with the use of an agate mortar and pestle until the size of the particles obtained was less than 44 $\mu \mathrm{m}$. For this, we sifted the powder through a mesh sieve. Figure~\ref{fig:incl} shows a scanning electron microscope (SEM) micrograph from the resulting powder that is very polydisperse as can be appreciated. This material was selected as the inclusion material due to its magnetic response, as the composite layers were subjected to a magnetic field in order to control their inclusion structure. Additionally, the magnetite powder was analyzed with a Panalytical Empyrean (Cu-k$_\alpha$, $\lambda = 1.5406 \, \mathring{\mathrm{A}}$) \hbox{X-ray} diffractometer to prove its purity. The diffractogram was compared with the Powder Diffraction File (PDF) with reference code 01-089-0691, showing a good correspondence for magnetite in its crystalline form (see figure 2 in reference \cite{cer14}).

\begin{figure}[t]
\begin{center}
\includegraphics[width=0.7\textwidth]{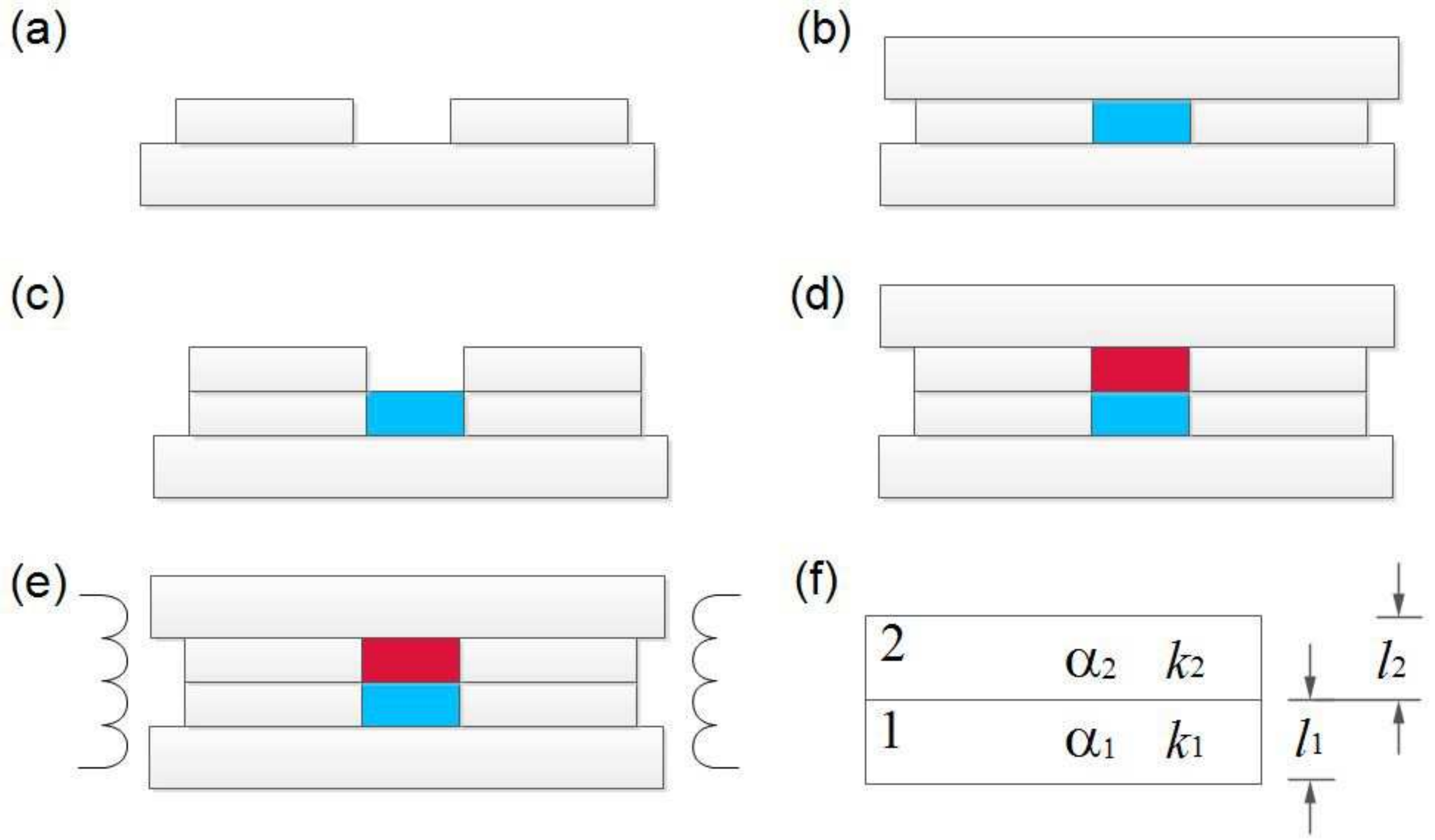} 
\caption{Schematic diagram of the steps followed in the sample preparation process. The layers were formed one by one to finally get a two-layer system. The blue square at the bottom in some of the figures corresponds to the first or substrate layer (layer 1), while the red square at the top to the second layer (layer 2) as identified in (f).}
\label{fig:samprep}
\end{center}
\end{figure}

The samples were prepared, layer by layer, on top of a microscope glass slide with cover slips (with thickness of about 200 $\mu \mathrm{m}$) piled up and glued together around the sample ``pocket'', in order to control the thickness of each layer (see figure \ref{fig:samprep}(a)). The mixture of resin (acting as the matrix) and magnetite particles (acting as the inclusions) was prepared in 1 ml syringes that had their tip cut off. For the first layer (layer 1) with thickness $l_1$, the mixture was spilt on top of the bottom glass slide and in between the cover slip piles. Afterwards, another glass slide was placed on top of the arrangement and the resin was let to cure as illustrated in figure \ref{fig:samprep}(b). In order to control the inclusion structure of the magnetite particles, the composite layers were subjected to a magnetic field with an intensity of 12.17 $\mathrm{kA} \, \mathrm{m}^{-1}$ during the polymerization process, applied through a pair of Helmholtz coils---to ensure the uniformity of the field---in the direction parallel to the surface of the layers (see figure \ref{fig:samprep}(e)). After 15 minutes, the top glass slide was removed from the arrangement and more cover slips added symmetrically on both sides of the sample to control the thickness of the second layer (see figure \ref{fig:samprep}(c)). By then, the first layer was solid enough to keep frozen the inclusion structure, but not completely cured so it would stick to the second layer to be applied. Following the same procedure, the second layer (layer 2) with thickness $l_2$ was formed on top of the first one. Figure \ref{fig:samprep} shows a step by step schematic diagram of the whole process. Special care was taken to avoid the formation of air bubbles inside the layers. This began with the preparation of the mixture itself, inside the syringe, working the plunger to remove the bubbles.

\begin{figure*}
\centering
\includegraphics[width=0.4\textwidth]{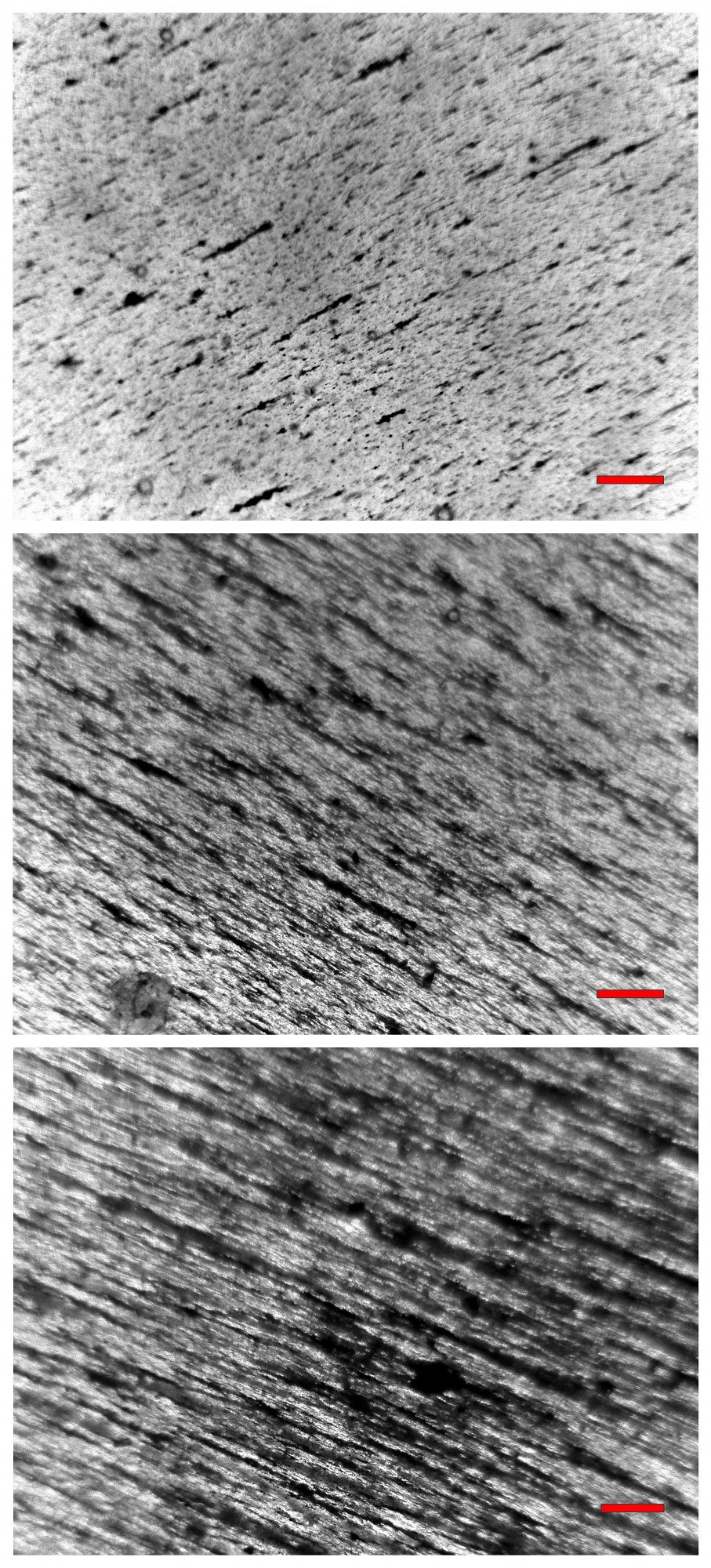}
\caption{Micrographs of the CSLs taken with an optical microscope with a magnification $\times$80. The red horizontal bar at the bottom right corner of each micrograph corresponds to a scale of 500 $\mu \mathrm{m}$. The concentration of inclusions changes from top to bottom as $\eta_\mathrm{m} = 0.005, 0.01, 0.02$.}
\label{fig:CSLs}
\end{figure*}

The concentration of magnetite particles, $\eta_{\rm m}$, was measured in volume fraction, considering that the magnetite used has a density of 5.2 g cm$^{-3}$. Two kinds of samples with total thickness around 850 $\mu {\rm m}$ were prepared: CCBs and RCBs, regarding the composition of the materials used for each layer. The RCBs consist in one side made of pure resin with thickness $l_1$ and the other made of composite material with thickness $l_2$, with a total thickness $l = l_1 + l_2$ (see figure \ref{fig:samprep}(f)). For these, the concentration of magnetite in the composite side was varied with $\eta_{\rm m} = 0.005, 0.01, 0.02$. Another control parameter was $x=l_2/l$, corresponding to the ratio of the thickness of the composite layer ($l_2$) and the total thickness of the sample ($l$). On the other hand, the CCBs consist in the same composite material on both sides, with the same concentration of inclusions and the same relative thickness, i.e., $l_1 = l_2$. For these, only the angle $\theta$ between the inclusion structures on each side of the interface was varied between 0$^\circ$ and 90$^\circ$ in steps of 22.5$^\circ$. After the curing process (about 8 hours), the samples were removed from the glass arrangement and cut in the form of an octagon of 9 mm in width.

\begin{figure*}
\centering
\includegraphics[width=\textwidth]{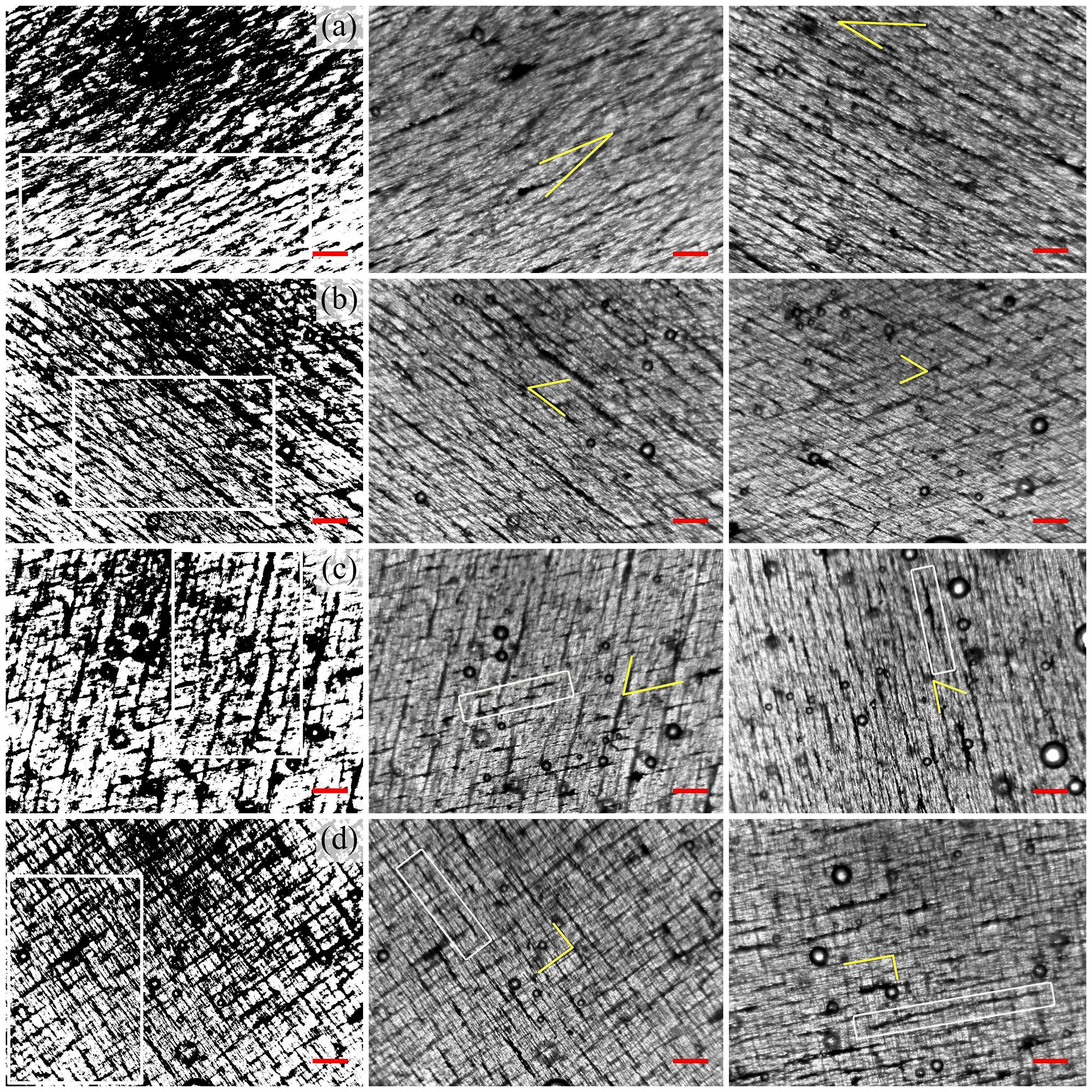}
\caption{Micrographs of CCBs with $\eta_\mathrm{m} = 0.01$, where the angle between the inclusion structures of the top and bottom layers increases from (a) to (d) as $\theta = 22.5^{\circ}, 45^{\circ}, 67.5^{\circ}, 90^{\circ}$, taken with an optical microscope with magnification $\times$80. The red horizontal bar at the bottom right corner of each micrograph corresponds to a scale of 500 $\mu \mathrm{m}$. The first column corresponds to binarized images obtained by processing the figures from the middle column. The middle column depicts the best focus of the interface between layers taken from the side of the top layer, i.e., layer 2. The last column depicts the best focus of the interface between layers taken from the side of the bottom layer, i.e., layer 1 or the \emph{substrate} layer. In the middle and last columns, the yellow lines depict the angle between the inclusion structures on each side. The thickness of the layers is the same on both sides $\sim 200 \, \mu \mathrm{m}$ (see text for more details).}
\label{fig:ImageJ}
\end{figure*}

In order to measure the thermal properties of the raw materials used in the preparation of the layered ones, samples consisting in a single layer with a thickness of approximately the same size as our two-layer systems were also prepared. In this way, one sample made of pure resin, one made of solid magnetite and tree composite single layers (CSLs) with $\eta_{\rm m} = 0.005, 0.01, 0.02$ were thermally characterized. \hbox{Figure \ref{fig:CSLs}} shows micrographs taken from CSLs with a thickness of $\sim 400 \, \mu \mathrm{m}$. The kind of inclusion structure formed by the arrangement of magnetite particles exhibits multifractal properties as discussed in \cite{cer14}, with the magnetite particles forming chain-like aggregates that grow with the concentration of inclusions as shown in the figure.

Additionally, with the objective to study the interface structure of the CCBs, equivalent bilayer samples of about 400 $\mu {\rm m}$ in total thickness were prepared, were the composite layers on each side had a thickness of about 200 $\mu \mathrm{m}$. The micrographs in middle column of figure \ref{fig:ImageJ} correspond to the best focus of the interface, taken from the side of the second layer (layer 2) while the last column corresponds to the best focus of the interface taken from the side of the first layer (the substrate layer or \hbox{layer 1}). The yellow lines are provided as an aid to identify the angle between the inclusion structures as seen from each side, while the red horizontal bar at the bottom right corner of each micrograph corresponds to a scale of 500 $\mu \mathrm{m}$. All of the micrographs were taken with an optical microscope with a magnification of $\times$80. The binarized images on the first column, obtained by processing the images in the middle one, were employed to characterize the interface structure via a lacunarity analysis as discussed later in \hbox{Section \ref{sec:resdis}}. We must mention that it was harder to control the formation of air bubbles in these samples, as each of their layers is very thin, about half the thickness of the layers present in the two-layer systems that were thermally characterized.

\begin{figure}[t]
\begin{center}
\includegraphics[width=0.6\textwidth]{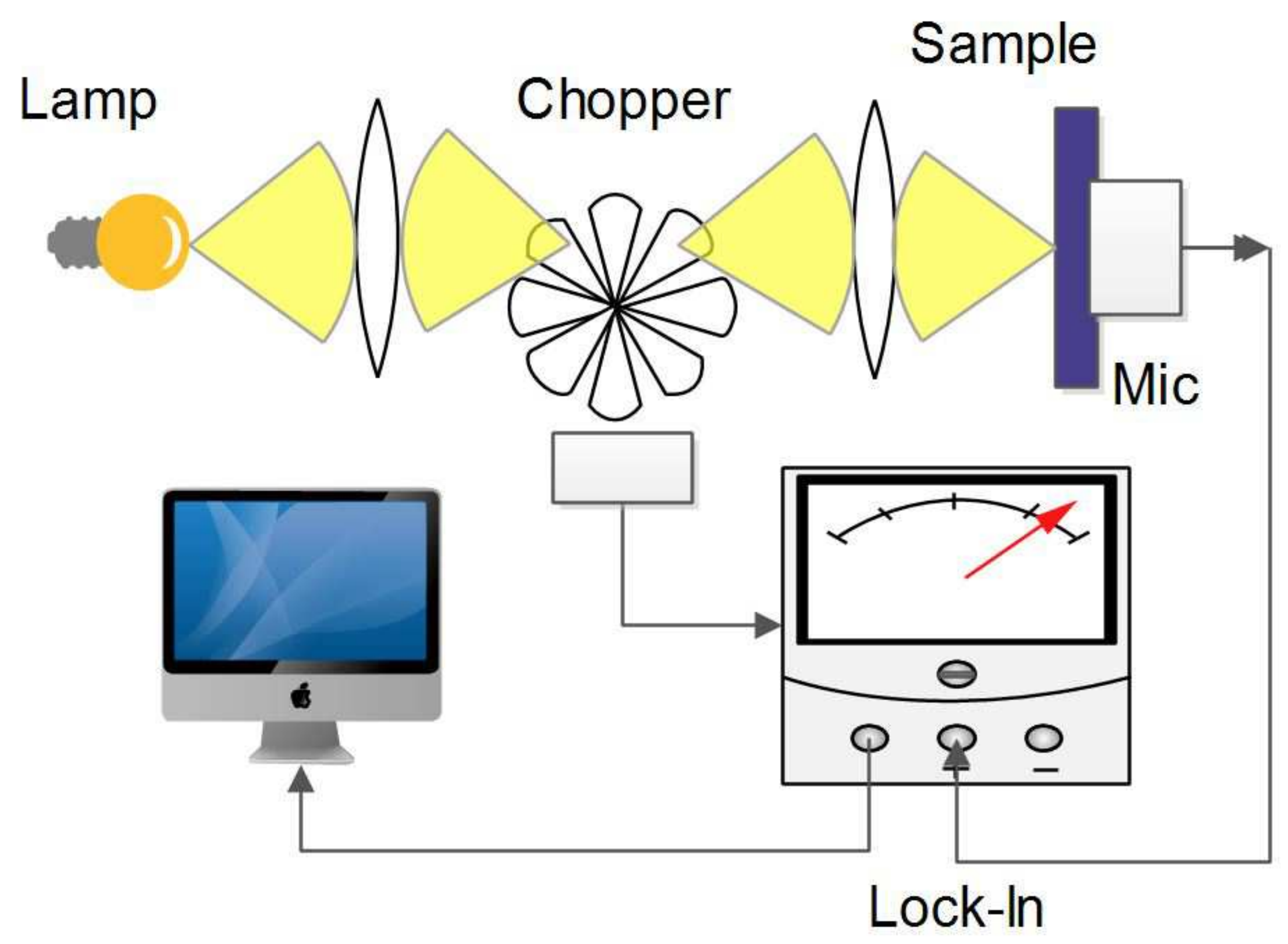} 
\caption{Schematic diagram of the experimental setup used to measure the thermal diffusivity $\alpha$, using the open-cell PA technique (see text for details).}
\label{fig:paset}
\end{center}
\end{figure}

\subsection{Experimental setups}
\label{ssec:exset}

In this work, we used the well established \emph{open-cell method} \cite{per87, fer89, mar91} in order to obtain the PA spectra of our samples. The experimental setup is schematically represented in the diagram of figure~\ref{fig:paset}. In this arrange\-ment, the sample is directly mounted onto a commercial electret microphone (RadioShack 270-0090). The beam of a 250 W tungsten lamp (HORIBA LSH-T250) was focused on the sample in such a way that the resulting beam had a diameter of about the same size as the inlet hole of the microphone \hbox{($\sim 3$ mm)}. The beam was then mechanically modulated with a Stanford Research Systems (SRS) optical chopper model SR540. As a result of the periodic heating of the sample by absorption of the modulated light, the microphone produces a signal that is fed to a lock-in amplifier (SRS model SR530). The illuminated face of the samples was painted with a matte black alkyd enamel in order to ensure that all of the light was absorbed in the surface of the sample. The paint coating amounts to about $20 \, \mu \mathrm{m}$ of the thickness of the samples. From the behaviour of the amplitude and phase of the PA signal, as a function of the modulation frequency, one can determine the thermal diffusivity $\alpha$ of a given sample.

\begin{figure}[t]
\begin{center}
\includegraphics[width=0.7\textwidth]{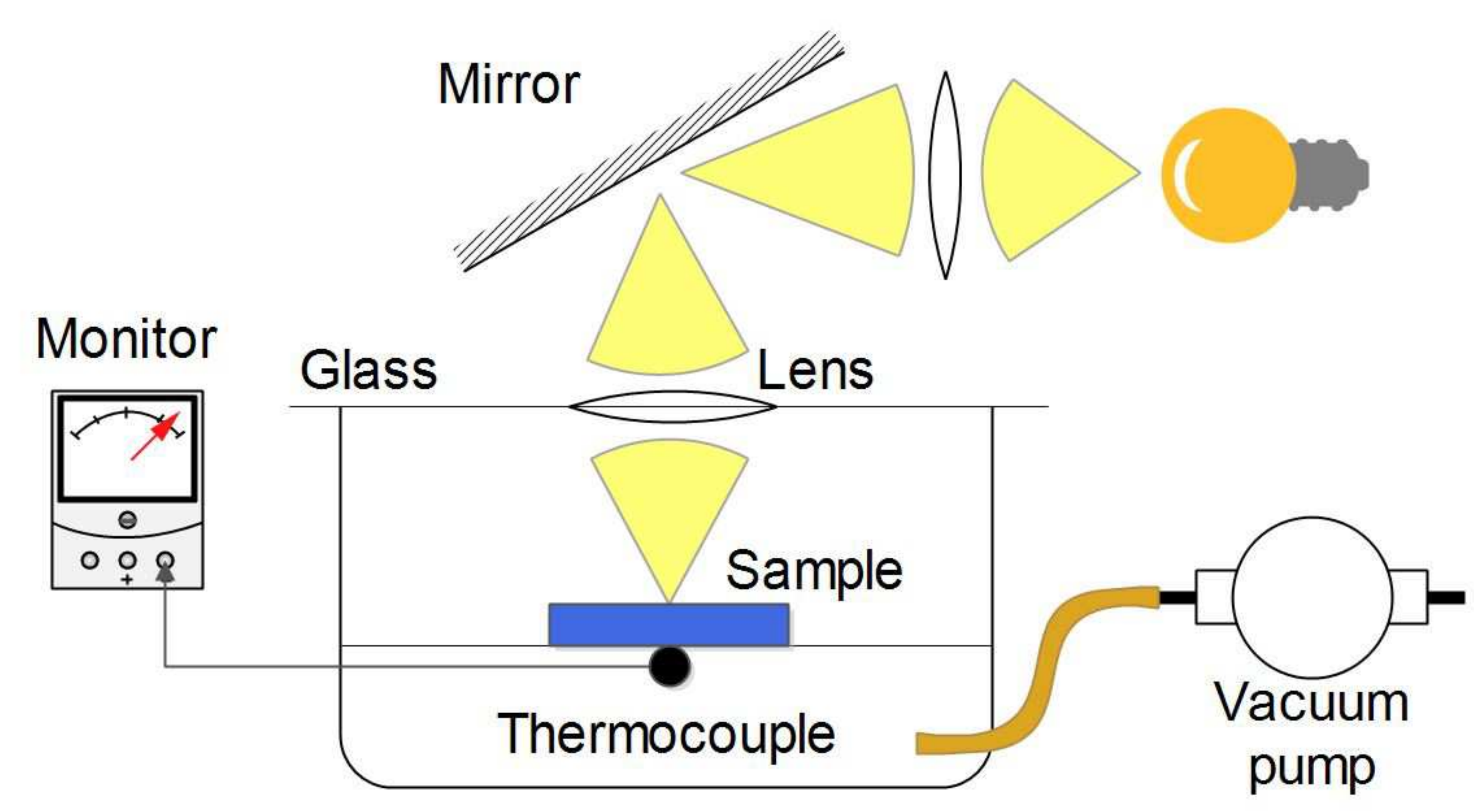} 
\caption{Schematic diagram of the experimental arrangement used to measure the volumetric heat capacity,  $\rho c$, using the thermal relaxation method (see text for details).}
\label{fig:trset}
\end{center}
\end{figure}

We also measured the volumetric heat capacity, $\rho c$, corres\-ponding to the product of the mass density and the constant pressure specific heat, respectively. For this, we employed the thermal relaxation method that is also well stablished in the literature \cite{hat79, gar06, val06}. Before the measurement, both faces of the sample are sprayed with the same matte black paint used before, in order to make its emissivity approximately equal to one. The sample is positioned inside a sealed chamber where a partial vacuum has been established. Then, the light beam from the 250 W lamp was focused with an arrangement of two lenses and a first surface mirror through a glass lid at the top of the chamber and onto one face of the sample. The temperature of the opposite non-illuminated face of the sample was then measured with Type K bead-wire temperature probe connected to a thermocouple monitor (Extech EA15). During the illumination process, the temperature of the sample rises to an equilibrium value above room temperature. Afterwards, the illumination is interrupted and its temperature traced as a function of time. The volumetric heat capacity ($\rho c$) can be calculated from the thermal decay that happens mainly through radiative processes. Figure~\ref{fig:trset} shows a schematic diagram of the experimental setup used for these measurements.

\begin{figure}[t]
\begin{center}
\includegraphics[width=0.55\textwidth]{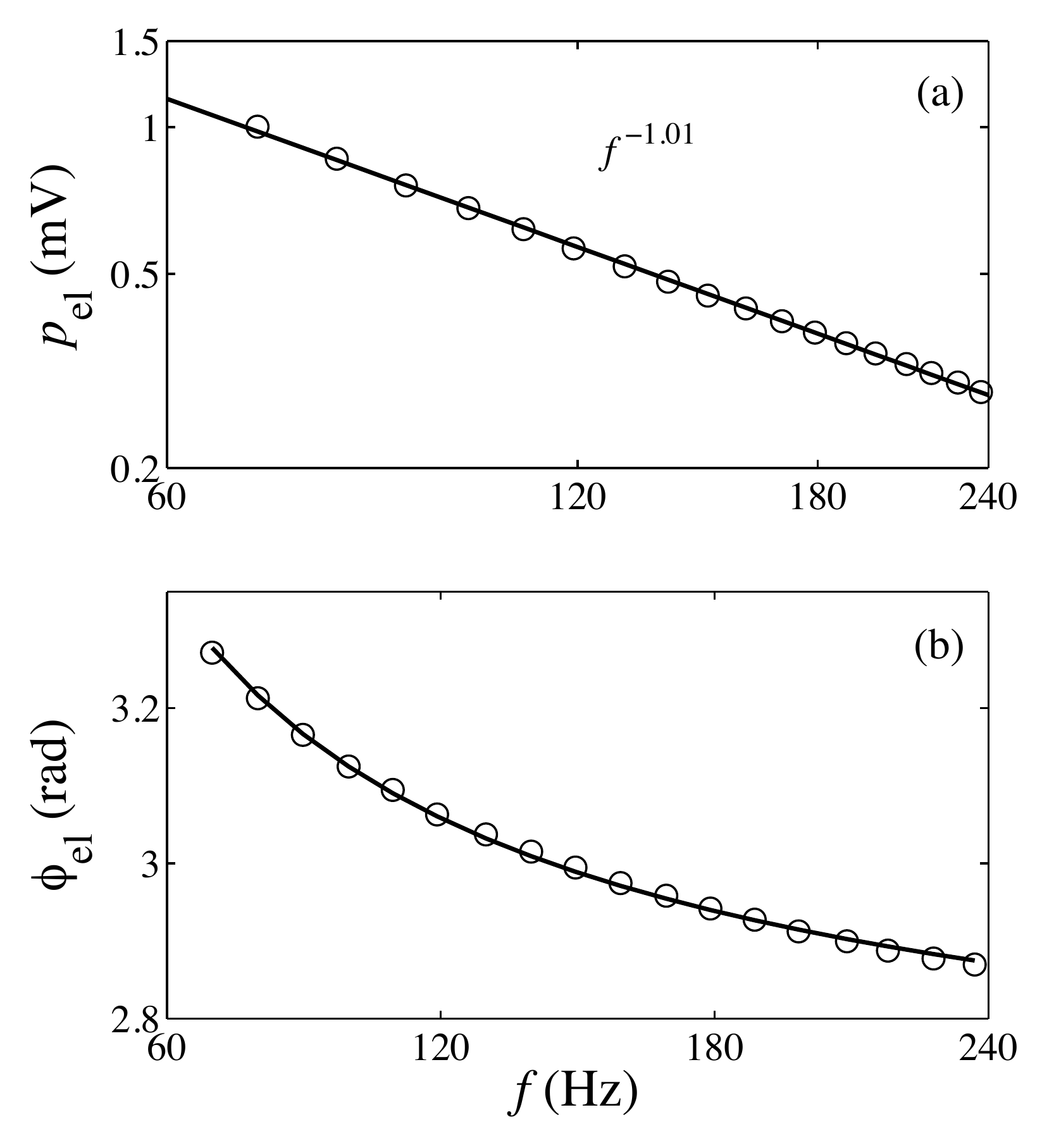}
\caption{Typical dependence of the PA signal's amplitude $p_\mathrm{el}$ (a) and phase $\phi_\mathrm{el}$ (b) on the modulation frequency $f$ for the samples studied in this work. The clear circles in the log-log plot in (a) correspond to the measured amplitude in mV, while the solid line corresponds to the fitted slope of the curve. In this case, a slope of $-1.01$ indicates that thermoelastic bending dominates in the generation of the PA signal. The clear circles in (b) correspond to the measured phase of the signal while the solid line corresponds to a fit with equation (\ref{eq:phiel}). The results shown correspond to the CCB with  $\theta = 90^{\circ}$ and $\eta_\mathrm{m} = 0.01$.}
\label{fig:pa-signal}
\end{center}
\end{figure}

\section{Results and discussion}
\label{sec:resdis}

\subsection{Determination of the thermal properties}
\label{ssec:tremprop}

The PA signal produced from platelike samples, when all of the light is absorbed at their surface, is known to have two main contributions: one coming from the \emph{thermal diffusion} phenomenon and the other from the \emph{thermoelastic bending} effect \cite{per87}. In this work we used the model from Rosencwaig and Gersho \cite{ros76}, that allows one to distinguish which one of these contributions dominates. The thermal diffusivity $\alpha$, can be obtained from the dependence of the PA signal's amplitude and phase on the modu\-lation frequency $f$, as discussed in details in \cite{dos02, per87}.

First, one must consider that a sample can be thermally thin or thick, regarding the ratio of its thermal diffusion length (that depends on the modulation frequency) and the thickness of the sample. These two regimes are separated by a cut-off frequency given by $f_\mathrm{c} = \alpha/(\pi l^2)$. Thermally thin samples fulfill the condition $f \ll f_\mathrm{c}$. In this regime, the amplitude of the PA signal behaves as $f^{-1.5}$, independent of the properties of the sample; this regime allows for the microphone's frequency response to be determined, that is later used to compensate the PA signal's amplitude. Conversely, thermally thick samples fulfill the condition $f \gg f_\mathrm{c}$. In this regime, if the thermal diffusion phenomenon dominates in the generation of the PA signal, its amplitude $p_\mathrm{td}$ and phase $\phi_\mathrm{td}$ depend on the modulation frequency as
\begin{eqnarray}
p_\mathrm{td} & = \frac{1}{f} \exp\left[ -\sqrt{b f} \right], 
\label{eq:ptd} \\
\phi_\mathrm{td} & = -\frac{\pi}{2} - \sqrt{b f}.
\label{eq:phitd}
\end{eqnarray}
On the other hand, if the thermoelastic bending contribution dominates, the amplitude $p_\mathrm{el}$ and the phase $\phi_\mathrm{el}$ of the PA signal depend on the modulation frequency as
\begin{eqnarray}
p_\mathrm{el} & \propto f^{-1}, \label{eq:pel} \\
\phi_\mathrm{el} & \simeq \frac{\pi}{2} + \arctan\left[ \frac{1}{\sqrt{b f} - 1} \right].
\label{eq:phiel}
\end{eqnarray}
In all of the cases, $b$ is the fitting parameter from which the thermal diffusivity can be estimated through the relation
\begin{equation}
\alpha = \frac{\pi l^2}{b}.
\label{eq:therm-diff}
\end{equation}

Figure~\ref{fig:pa-signal} shows the typical dependence of the amplitude and phase of the PA signal on the modulation frequency for the two-layer samples we studied in this work. The results shown in the figure correspond the CCB with $\theta = 90^{\circ}$ and $\eta_\mathrm{m} = 0.01$. A slope of $-1.01$ shown by $p_\mathrm{el}$ as a function of $f$ in the log-log plot of figure~\ref{fig:pa-signal}(a), means that the thermoelastic bending effect dominates in the generation of the PA signal. The value of the thermal diffusivity $\alpha$ was then obtained by fitting the experimental data for the phase of the PA signal with equation (\ref{eq:phiel}), shown in figure~\ref{fig:pa-signal}(b). All of our two-layer samples, including the pure resin one, exhibited this behaviour. Only the pure solid magnetite sample showed thermal diffusion as the dominating PA effect. The experimental data for this sample were fitted instead with equations (\ref{eq:ptd}) and (\ref{eq:phitd}).

\begin{figure}[t]
\begin{center}
\includegraphics[width=0.7\textwidth]{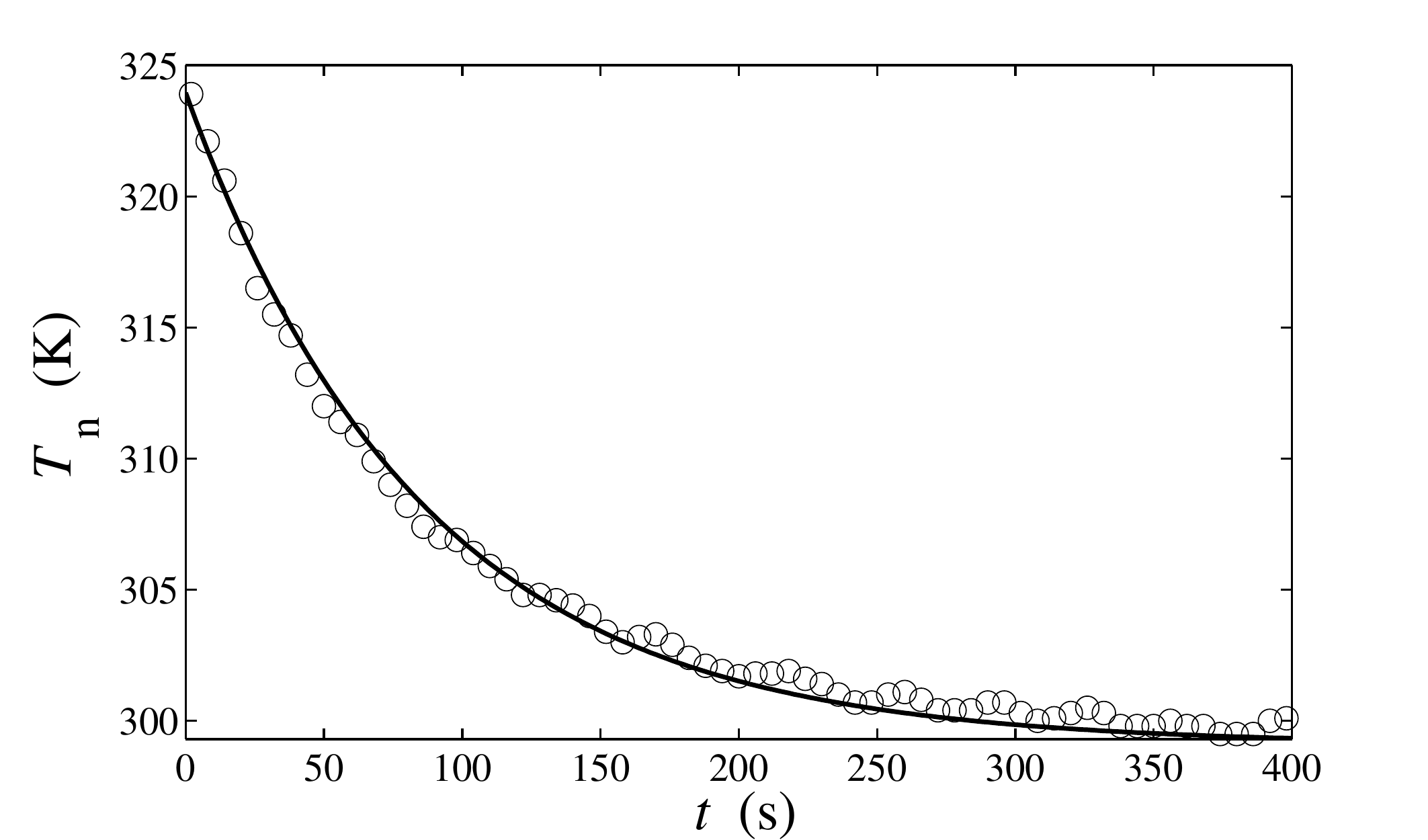} 
\caption{Typical experiment of thermal relaxation. The clear circles correspond to the measured  temperature $T_\mathrm{n}$ of the non-illuminated face as a function of the time $t$ for the pure resin sample with thickness $l = 829 \, \mu {\rm m}$. The solid line corresponds to a fit with \hbox{equation (\ref{eq:thermdecay})}.}
\label{fig:therm-decay}
\end{center}
\end{figure}

In order to determine the volumetric heat capacity $\rho c$, we used the thermal relaxation method. For this technique, one face of the sample is illuminated  with a constant flux of light as illustrated in the schematic diagram shown in figure~\ref{fig:trset}. As a consequence, a lack of equilibrium between the illuminated and non-illuminated faces of the sample is established. This phenomenon can be approximately described by a 1D equation when the thickness $l$ of the sample is much smaller than its width. The energy conservation condition is given by
\begin{equation}
I_0 - \sigma T_\mathrm{i}^4 - \sigma T_\mathrm{n}^4 = \frac{\mathrm{d}}{\mathrm{d}t} \int_0^{l_\mathrm{s}} \rho c T(x,t) \, \mathrm{d}x,
\label{eq:engcon1}
\end{equation}
where $I_0$ is the flux of incident light, $\sigma$ is the Stefan-Boltzmann constant, $T_\mathrm{i}$ is the temperature of the \emph{illuminated} face, $T_\mathrm{n}$ is the temperature of the \emph{non-illuminated} face, $\rho$ is the mass density of the sample and $c$ is its specific heat at constant pressure. Here, we have explicitly used the fact that both faces of the sample are coated with a thin layer of black paint that has an emissivity coefficient approximately equal to one \cite{hat79}.

For long times, when thermal equilibrium is established, and for the values of $l$ and $I_0$ used, the condition $l \, \mathrm{d}T(x,t)/\mathrm{d}x \ll T_\mathrm{i}(t) \approx T_\mathrm{n}(t)$ is fulfilled. Moreover, using the fact that $c$ does not depend on the position and that it is practically constant in the interval of a few degrees above room temperature, equation (\ref{eq:engcon1}) can be solved for the thermal decay of the non-illuminated face after the illumination on the sample is interrupted. Considering this process is mainly due to radiative processes, the solution can be written as
\begin{equation}
T_\mathrm{n}(t) = T_\mathrm{n,0} + (T_\mathrm{n,\,max} - T_\mathrm{n,0}) \exp(-t/\tau_\mathrm{d}),
\label{eq:thermdecay}
\end{equation}
where $T_\mathrm{n,0}$ is the final temperature reached by the non-illuminated face after cooling down, while $T_\mathrm{n,\,max}$ is its maximum saturation temperature reached before the illumination is interrupted. The relaxation mean time $\tau_\mathrm{d}$ is related to $\rho c$ by
\begin{equation}
\tau_\mathrm{d} = \frac{\rho c l}{8 \sigma T_\mathrm{n,0}^3}.
\label{eq:tau}
\end{equation}
Details for the solution of equation (\ref{eq:engcon1}) can be found in \cite{dos02}. Afterwards, the thermal conductivity $k$ can be calculated from the equation
\begin{equation}
k = \rho c \alpha.
\label{eq:thermcond}
\end{equation}
A summary of the results for the samples studied in this work is presented in table~\ref{tab:tab1}.

\begin{table*}[t]
\centering
\caption{\label{tab:tab1} Summary of the results for the properties of the samples studied in this work. The second block corresponds to the CSLs. The third and fourth blocks correspond to the RCBs and the CCBs, respectively.}
 
\begin{indented}
\lineup
\item[]\begin{tabular}{@{}lllllll}
\br
$\eta_\mathrm{m}$ (v.f.) & $l$ $(\mu\mathrm{m})$ & $x$ & $\theta$ (deg) & $\alpha \times 10^{-6}$ $(\mathrm{m}^2 \, \mathrm{s}^{-1})$ & $k$ (W m$^{-1}$ K$^{-1}$) \cr
\mr
Resin & 829 & & & 20.50 $\pm$ 0.60 & 23.10 $\pm$ 0.69 \cr
Magnetite & 899 & & & 48.30 $\pm$ 0.07 & 68.10 $\pm$ 0.67 \cr
\mr
0.005 & 790 & & & 33.00 $\pm$ 1.84 & 29.99 $\pm$ 1.74 \cr
0.010 & 870 & & & 39.60 $\pm$ 1.69 & 45.39 $\pm$ 2.01 \cr
0.020 & 823 & & & 30.00 $\pm$ 0.97 & 46.67 $\pm$ 1.58 \cr
\mr
0.010 & 850 & 0.25 & & 22.90 $\pm$ 0.01 & 21.74 $\pm$ 0.31 \cr
0.010 & 900 & 0.50 & & 30.00 $\pm$ 1.15 & 32.31 $\pm$ 1.30 \cr
0.010 & 840 & 0.75 & & 38.50 $\pm$ 1.40 & 42.63 $\pm$ 1.65 \cr
\mr
0.005 & 896 & 0.50 & 22.5$^{\circ}$ & 33.90 $\pm$ 0.88 & 32.18 $\pm$ 0.94 \cr
0.010 & 842 & 0.50 & 22.5$^{\circ}$ & 38.00 $\pm$ 1.65 & 38.39 $\pm$ 1.76 \cr
0.020 & 844 & 0.50 & 22.5$^{\circ}$ & 36.60 $\pm$ 1.85 & 45.48 $\pm$ 2.37 \cr
0.010 & 793 & 0.50 & 45.0$^{\circ}$ & 23.80 $\pm$ 1.26 & 27.28 $\pm$ 1.49 \cr
0.010 & 840 & 0.50 & 67.5$^{\circ}$ & 34.60 $\pm$ 0.83 & 42.05 $\pm$ 1.14 \cr
0.010 & 906 & 0.50 & 90.0$^{\circ}$ & 41.00 $\pm$ 1.42 & 48.04 $\pm$ 1.76 \cr
\br
\end{tabular}
\end{indented}
\end{table*}

Figure~\ref{fig:therm-decay} shows a typical experiment of thermal relaxation, in this case, for the pure resin sample. The clear circles correspond to the measured temperature $T_\mathrm{n}$ as a function of time, while the solid line was obtained by fitting equation (\ref{eq:thermdecay}) to the experimental data with $\tau_\mathrm{d}$ as the fitting parameter. We must mention that we tested our experimental setups by measuring the thermal properties of a \emph{p}-type silicon sample with thickness $l = 535 \, \mu\mathrm{m}$. For this sample, the thermal diffusion phenomenon dominates in the generation of the PA signal. The thermal diffusivity and conductivity obtained from our experimental results are $87.7 \pm 3.3 \times 10^{-6} \, \mathrm{m}^2 \, \mathrm{s}^{-1}$ and $143.4 \pm 5.7 \, \mathrm{W} \, \mathrm{m}^{-1} \, \mathrm{K}^{-1}$, respectively, that agree with the reported values $87  \times 10^{-6} \, \mathrm{m}^2 \, \mathrm{s}^{-1}$ for the thermal diffusivity \cite{per87} and $149 \, \mathrm{W} \, \mathrm{m}^{-1} \, \mathrm{K}^{-1}$ for the thermal contuctivity \cite{crc05} within error values.

\begin{figure}[tb]
\begin{center}
\includegraphics[width=0.57\textwidth]{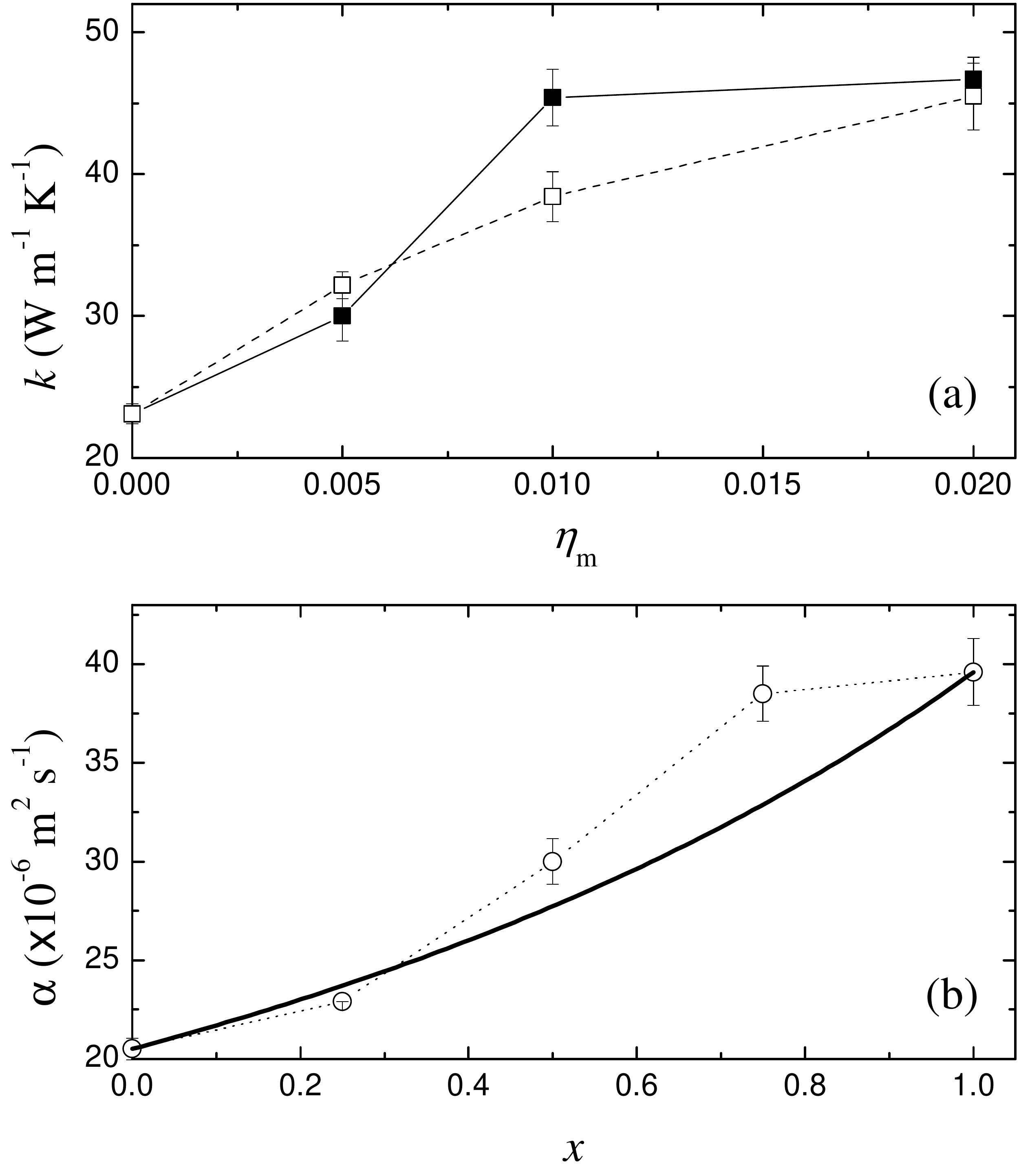}
\caption{(a) Measured thermal conductivity of the CSLs (solid curve with solid squares) and the effective thermal conductivity of the CCBs with $\theta = 22.5^{\circ}$ (dashed curve with clear squares) as a function of $\eta_\mathrm{m}$. The values obtained for the CSLs are equivalent to those obtained for the CCBs with $\theta = 0^{\circ}$. In (b) the dotted curve with clear circles corresponds to the measured effective thermal diffusivity $\alpha$ as a function of $x = l_2/l$ for the RCBs. In this case, the composite side has a concentration of inclusions $\eta_\mathrm{m} = 0.01$. The thick solid curve corresponds to a plot of equation (\ref{eq:eff-diff}) calculated with the values obtained for the corresponding CSL and the pure resin sample.}
\label{fig:k-alpha}
\end{center}
\end{figure}

\subsection{Interface thermal resistance}
\label{ssec:int-therm_res}

First of all, let us analyze the properties of the CSLs. As we showed in \cite{cer14}, the inclusion structure formed by the chains of magnetite particles in this kind of anisotropic samples exhibits multifractal properties. Previously, we prepared them in bulk and platelike samples were later extracted from the center of a $\sim 3 \, \mathrm{cm}^{3}$ cube in order to avoid surface effects from the mold. The measured thermal conductivity of samples with similar inclusion structure as our CSLs, i.e., with the magnetite chains parallel to the faces of the samples, was smaller that that of the resin and the magnetite themselves. We associated this result with the overlap of resin and magnetite domains---by domains meaning areas occupied by either resin or magnetite aggregates---from different substrates of the samples, increasing the thermal resistance in the direction of transmission of the heat, as the layers of inclusion structure were not aligned in this direction. Here, our CLSs were prepared in platelike form between glass slides, with the inclusion structure in close reach of the slides' surfaces. The measured thermal conductivity for these samples is higher that of the resin but smaller than the value obtained for the magnetite. Clearly, there must be a strong surface effect that allows the magnetite chains to also develop vertically, i.e., perpendicular to the faces of the samples. 

Figure \ref{fig:CSLs} shows micrographs from our CSLs with a thickness of about \hbox{400 $\mu$m}, that is very close to the thickness of the layers found in the CCBs, for example. As can be appreciated, not much overlap between domains of resin and magnetite is noticeable from different substrates of the samples. This means that the magnetite chains are able to stretch from one face to the other inside the sample, creating pathways for the heat to flow, consequently enhancing the thermal properties of our CLSs with respect to their resin matrix. Figure \ref{fig:k-alpha}(a) shows the results for the measured thermal conductivity of these samples as a function $\eta_\mathrm{m}$. For the concentrations of magnetite used in this work, the thermal conductivity increases with $\eta_\mathrm{m}$ as shown by the solid curve with solid squares. This provided us with a good starting point, as our composite layers had a larger thermal conductivity than that of the resin. It is worth to mention that this kind of enhancement in the thermal properties, due to the anisotropic formation of chain-like aggregates of particles with higher conductivity than the supporting material, has also been observed in magnetic fluids under the action of applied fields \cite{li05, phi07, fan09}.

Regarding our two-layer systems, we will first discuss the RCBs. For these, we only considered samples where the composite layer has a concentration $\eta_\mathrm{m} = 0.01$. These samples consist in a resin layer and a composite one, with thicknesses around 200, 400 or \hbox{600 $\mu$m} depending on the value of $x \, (=l_2/l)$. We must point out that, for the modulation frequencies used in this work, the layers with thicknesses around $200 \, \mu\mathrm{m}$ (either resin or composite) are thermally thin with a cut-off frequency $f_c \geq 163 \, \mathrm{Hz}$. Layers with thicknesses larger than 400 $\mu$m are already thermally thick for the modulation frequencies used. Measuring the thermal diffusivity of bilayers consisting in layers with different thermal thicknesses requires a special care, as it is known that their thermal properties tend to the values of the thermally thick layer \cite{mar96}. In this way, the PA spectrum of RCBs with layers of different thermal thicknesses were measured with their thermally thin layer facing the modulated light beam. 

On the other hand, the effective thermal diffusivity of two-layer systems can be described by the equation
\begin{equation}
\alpha = \left[ \frac{x^2}{\alpha_2} + \frac{(1-x)^2}{\alpha_1} + \frac{2x(1-x)}{\sqrt{\alpha_1 \alpha_2}} \right]^{-1},
\label{eq:eff-diff}
\end{equation}
when both layers are thermally thick or, as a particular case, when their thermal effusivities are equal regardless of their thermal thicknesses \cite{man90, mar96}. For the case of our RCBs, $\alpha_1$ corresponds to the thermal diffusivity of the pure resin layer with thickness $l_1$, $\alpha_2$ to the composite layer with thickness $l_2$, and $x=l_2/l$ with $l = l_1 + l_2$ as defined before. The dotted curve with clear circles in figure \ref{fig:k-alpha}(b) corresponds to the measured thermal diffusivity of these samples, as a function of $x$, while the thick solid line to equation (\ref{eq:eff-diff}), calculated using the values obtained for the pure resin sample (the value at $x=0$) and the CSL with $\eta_\mathrm{m} = 0.01$ (the value at $x=1$). Considering that some of our RCBs have layers with different thermal thicknesses, the good correspondence between our experimental data with this approximation suggests that the resin and composite layers have almost identical values for their effusivities.

\begin{figure}[tb]
\begin{center}
\includegraphics[width=0.65\textwidth]{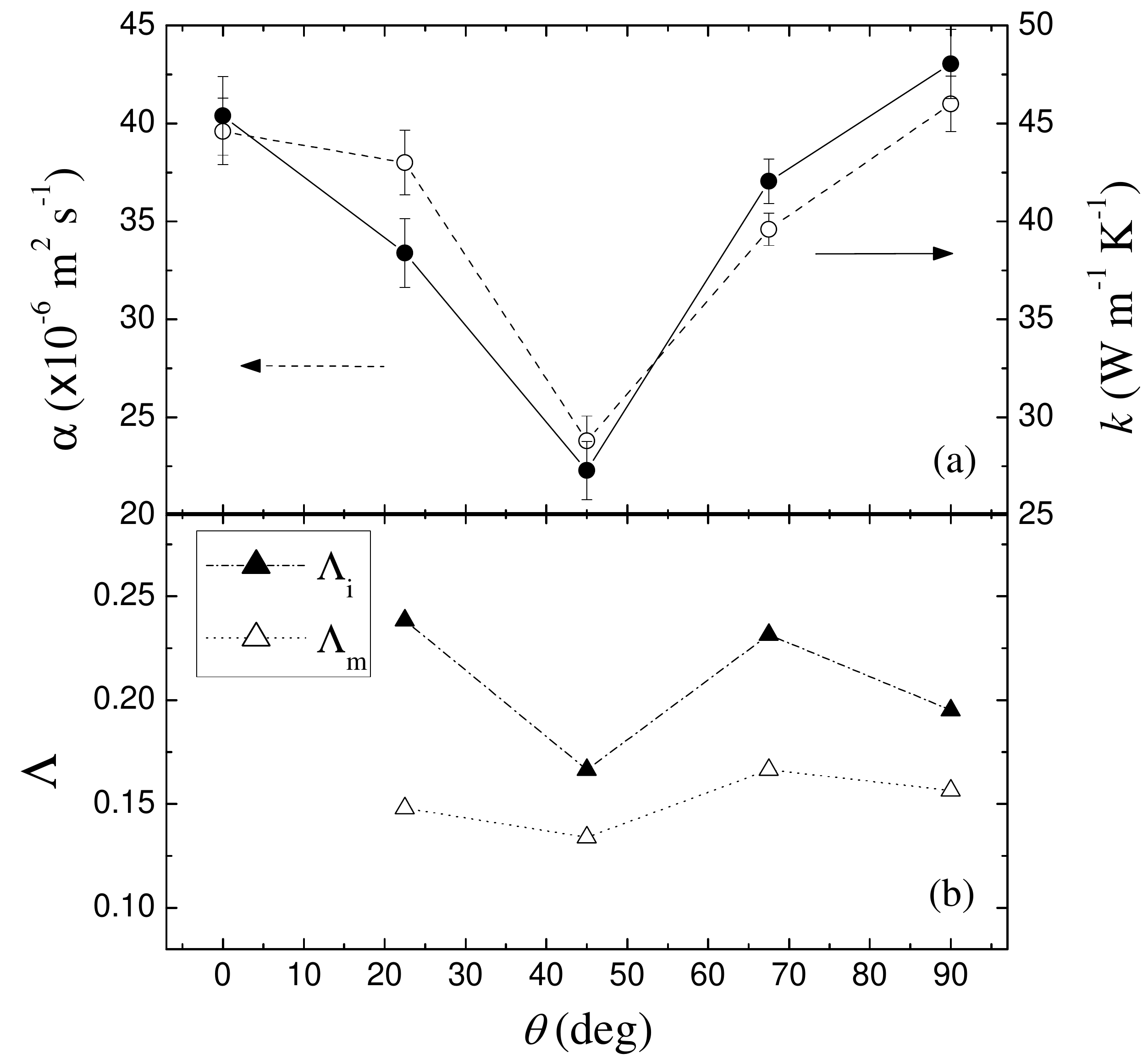}
\caption{(a) Effective thermal diffusivity $\alpha$ (left axis and dashed curve with clear circles) and conductivity $k$ (right axis and solid curve with solid circles) as a function of the angle $\theta$ for CCBs with $\eta_\mathrm{m} = 0.01$. (b) Mean lacunarity $\Lambda$ for the \emph{interface structure} of the samples in (a) as a function of the angle $\theta$. The dashed curve with solid triangles corresponds to the mean lacunarity of the inclusions, $\Lambda_\mathrm{i}$, while the dotted curve with clear triangles corresponds to the mean lacunarity of the matrix, $\Lambda_\mathrm{m}$ (see text for more details).}
\label{fig:k-lac}
\end{center}
\end{figure}

We will now focus our attention on the CCBs. In these symmetric bilayers, only the angle between the inclusion structures on each side of the interface changes. Results for the measured thermal diffusivity $\alpha$ of CCBs with $\eta_\mathrm{m} = 0.01$ are presented in figure \ref{fig:k-lac}(a), with the left axis and the dashed curve with clear circles. Notice how $\alpha$ first decreases with the increasing $\theta$, reaching a minimum at $\theta = 45^{\circ}$, to later increase again. Equation (\ref{eq:eff-diff}) predicts a constant value for the effective thermal diffusivity of our CCBs, given the fact that both of their composite layers are thermally thick. The nontrivial dependance of $\alpha$ on $\theta$ means that there is a thermal resistance with origin at the interface. This is an emergent effect resulting from the overlap of inclusion structures in this region, not exhibited by the RCBs. Moreover, the volumetric heat capacity $\rho c$ is practically the same for all our CCBs, so the V-shaped dependance of $\alpha$ on $\theta$ is also reflected on their thermal conductivity $k$, corresponding to the right axis and the solid curve with solid circles in figure \ref{fig:k-lac}(a). In order to explain these results, let us analyze the inclusion structures of CCBs on both sides of the interface for the different orientations. 

The last column of figure \ref{fig:ImageJ} shows the best focus on the interface taken from the side of layer 1 as explained before. Notice how the magnetite aggregates in these layers, form long and thick chains. This kind of structures, pointed out with the white rectangles for $\theta = 67.5^{\circ}, 90^{\circ}$, exposes a large effective area to the transmission of heat due to their fractal properties. These kind of structures can also be seen in \hbox{layer 2} for $\theta \leq 45^{\circ}$ in the middle column. Nonetheless, for $\theta > 45^{\circ}$, the magnetite aggregates seen in \hbox{layer 2} are thinner and shorter as pointed out with the white rectangles for $\theta = 67.5^{\circ}, 90^{\circ}$ in the same column. This kind of aggregates exposes less effective area to the transmission of heat but tend to be denser, developing vertically from face to face inside the layer. Their shape comes about the presence of the magnetized aggregates of layer 1, forming a close to right angle with the direction of the applied field during their formation. In summary, the inclusion structure of magnetite particles in layer 2 for $\theta > 45^{\circ}$ is very different from the inclusions structures of the other layers in our CCBs, either layer 1 or layer 2.

In this way, the thermal response shown in figure \ref{fig:k-lac}(a) can be explained from the point of view of two competing pathways for heat transport at the interface of our CCBs: (i) the chain-to-chain one and (ii) the one between magnetite and resin domains. For $\theta$ between 0$^{\circ}$ and 45$^{\circ}$, the purposefully induced overlap of resin and magnetite domains between the inclusion structures of each layer at the interface confirms our previous findings \cite{cer14}, as the thermal resistance of the interface increases with $\theta$. Indeed, it was this result that inspired us to study two-layer systems in order to investigate this effect, in a more controlled manner, by varying the angle between anisotropic inclusion structures in order to expose a larger effective area between domains. This overlap becomes maximal for $\theta = 45^{\circ}$. For $\theta > 45^{\circ}$, an interesting effect form the magnetized substrate of layer 1, renders the magnetite chains of layer 2 thiner and shorter as $\theta$ increases, decreasing the overlap of resin and magnetite domains at the interface and favouring the chain-to-chain heat transport again.


In order to quantitatively characterize the \emph{interface structure} of our CCBs, the mean lacunarity of the inclusion ($\Lambda_\mathrm{i}$) and of the matrix ($\Lambda_\mathrm{m}$) patterns were obtained by analyzing the binarized images presented in the first column of figure~\ref{fig:ImageJ} with the plugin \emph{FracLac} for \emph{ImageJ}. The regions studied are enclosed by white rectangles in those images. These regions were selected as representative of their corresponding interface structure for the balance of inclusions and resin shown in their binarized form. The results are presented in figure~\ref{fig:k-lac}(b). It is worth to mention that the concept of lacunarity was first introduced by Mandelbrot \cite{man83} and has undergone several improvements by taking into account the set of scales involved in the iteration process \cite{all91, aar94}, to become a multiscaled method for describing patterns of spatial dispersion \cite{plo96}. The mean lacunarity ($\Lambda$) indicates how the space is filled by describing the distribution of the sizes of gaps or \emph{lacunae} in a given structure. Greater lacunarity reflects a greater size distribution of the lacunae or a higher degree of ``gappiness''. From this point of view, textured patterns can be studied regarding their inhomogeneity and translational and rotational invariance. Lacunarity analyses have been performed for the study of very different systems \cite{mon11, gou11}.

For the interface structure of our CCBs, this means that the larger $\Lambda$ is the less overlap between resin and magnetite domains. As expected, the dependance of the mean lacunarity for the inclusion and the resin structures, at the interface, resembles the V-shaped thermal response observed for $\alpha$ and $k$ as a function of $\theta$. This is sown in figure \ref{fig:k-lac}(b), where one can appreciate that both $\Lambda_\mathrm{i}$ and $\Lambda_\mathrm{m}$ also reach a minimum at $\theta = 45^{\circ}$. Moreover, one can expect a similar thermal response, as a function of $\theta$, if the concentration of inclusions changes in the CCBs. As shown by the dashed curve with clear squares in figure \ref{fig:k-alpha}(a), the thermal conductivity of CCBs with $\theta = 22.5^{\circ}$ is proportional to $\eta_\mathrm{m}$. From this, the thermal response presented in figure \ref{fig:k-lac}(a) could, in principle, be shifted up or down by varying $\eta_\mathrm{m}$ as long as it remains low.

\section{Conclusions}
\label{conc}

In this paper, we have studied the thermal properties of two-layer systems consisting in polyester resin layers and composite layers with an anisotropic inclusion structure of magnetite particles. The thermal properties of bilayers consisting in a resin layer and a composite layer can be described by well established models. On the other hand, bilayers consisting in two composite layers, where only the angle between the anisotropic inclusion structures on the sides of the interface is varied, show a nontrivial dependance of their thermal properties on this angle. We were able to explain this from the point of view of two competing pathways for heat transport: one coming from the close contact of magnetite chains and the other from the overlap of resin and magnetite domains at the interface of these bilayers. The study presented here confirms our previous findings regarding the latter and exhibits new ways to develop composite layered materials with a controllable thermal response. Additionally, a new way for controlling the formation of complex inclusion structures in composite layers is suggested, with the use of magnetized substrates that can later be removed. Our results can be used in the development of intelligent materials with a real-time controllable thermal response. We are pursuing this line of research, along with theoretical investigations that correlate the complexity of the inclusion structures with the nontrivial thermal response observed in this kind of composite systems.

\ack

This work was partially supported by CONACyT, by SEP through the grant PROMEP/103.5/10/7296, and by VIEP-BUAP through the grants DORV-EXC13-I and CAEJ-EXC14-G. The authors are grateful to R Silva-Gonz\'alez (BUAP) for the SEM micrographs and to M E Mendoza (BUAP) for the optical ones.

\end{document}